\documentclass[11pt]{article}

\usepackage{latexsym,amsmath,amssymb}
\usepackage{graphicx}

\newcounter{subequation}[equation]
\makeatletter

\def\bcite{\@ifnextchar [{\@tempswatrue\@bcitex}{\@tempswafalse\@bcitex[]}}
\def\@bcitex[#1]#2{\if@filesw\immediate\write\@auxout{\string\citation{#2}}\fi
  \let\@bcitea\@empty
  \@bcite{\@for\@bciteb:=#2\do
    {\@bcitea\def\@bcitea{,\penalty\@m\ }%
     \def\@tempa##1##2\@nil{\edef\@bciteb{\if##1\space##2\else##1##2\fi}}%
     \expandafter\@tempa\@bciteb\@nil
     \@ifundefined{b@\@bciteb}{{\reset@font\bf ?}\@warning
       {Citation `\@bciteb' on page \thepage \space undefined}}%
     \hbox{\csname b@\@bciteb\endcsname}}}{#1}}
\def\@bcite#1#2{{#1\if@tempswa , #2\fi}}

\def\thesubequation{\theequation\@alph\c@subequation}
\def\@subeqnnum{{\rm (\thesubequation)}}
\def\slabel#1{\@bsphack\if@filesw {\let\thepage\relax
   \xdef\@gtempa{\write\@auxout{\string
      \newlabel{#1}{{\thesubequation}{\thepage}}}}}\@gtempa
   \if@nobreak \ifvmode\nobreak\fi\fi\fi\@esphack}
\def\subeqnarray{\stepcounter{equation}
\let\@currentlabel=\theequation\global\c@subequation\@ne
\global\@eqnswtrue
\global\@eqcnt\z@\tabskip\@centering\let\\=\@subeqncr
$$\halign to \displaywidth\bgroup\@eqnsel\hskip\@centering
  $\displaystyle\tabskip\z@{##}$&\global\@eqcnt\@ne
  \hskip 2\arraycolsep \hfil${##}$\hfil
  &\global\@eqcnt\tw@ \hskip 2\arraycolsep
  $\displaystyle\tabskip\z@{##}$\hfil
   \tabskip\@centering&\llap{##}\tabskip\z@\cr}
\def\endsubeqnarray{\@@subeqncr\egroup
                     $$\global\@ignoretrue}
\def\@subeqncr{{\ifnum0=`}\fi\@ifstar{\global\@eqpen\@M
    \@ysubeqncr}{\global\@eqpen\interdisplaylinepenalty \@ysubeqncr}}
\def\@ysubeqncr{\@ifnextchar [{\@xsubeqncr}{\@xsubeqncr[\z@]}}
\def\@xsubeqncr[#1]{\ifnum0=`{\fi}\@@subeqncr
   \noalign{\penalty\@eqpen\vskip\jot\vskip #1\relax}}
\def\@@subeqncr{\let\@tempa\relax
    \ifcase\@eqcnt \def\@tempa{& & &}\or \def\@tempa{& &}
      \else \def\@tempa{&}\fi
     \@tempa \if@eqnsw\@subeqnnum\refstepcounter{subequation}\fi
     \global\@eqnswtrue\global\@eqcnt\z@\cr}
\let\@ssubeqncr=\@subeqncr
\@namedef{subeqnarray*}{\def\@subeqncr{\nonumber\@ssubeqncr}\subeqnarray}
\@namedef{endsubeqnarray*}{\global\advance\c@equation\m@ne%
                           \nonumber\endsubeqnarray}

\renewcommand\maketitle{\par
  \begingroup
    \if@twocolumn
      \ifnum \col@number=\@ne
        \@maketitle
      \else
        \twocolumn[\@maketitle]%
      \fi
    \else
      \newpage
      \global\@topnum\z@   
      \@maketitle
    \fi
    \thispagestyle{plain}\@thanks
  \endgroup
  \setcounter{footnote}{0}%
  \global\let\thanks\relax
  \global\let\maketitle\relax
  \global\let\@maketitle\relax
  \global\let\@thanks\@empty
  \global\let\@author\@empty
  \global\let\@date\@empty
  \global\let\@title\@empty
  \global\let\title\relax
  \global\let\author\relax
  \global\let\date\relax
  \global\let\and\relax
}

\makeatother

\DeclareFontFamily{OT1}{rsfs10}{}
\DeclareFontShape{OT1}{rsfs10}{m}{n}{ <-> rsfs10 }{}
\DeclareMathAlphabet{\mathscript}{OT1}{rsfs10}{m}{n}

\numberwithin{equation}{section}

\newcommand{\ns}{\normalsize}
\newcommand{\pt}{\partial}

\newcommand{\be}{\begin{equation}}
\newcommand{\ee}{\end{equation}}
\newcommand{\nn}{\nonumber}
\newcommand{\bea}{\begin{eqnarray}}
\newcommand{\eea}{\end{eqnarray}}
\newcommand{\bsea}{\begin{subeqnarray}}
\newcommand{\esea}{\end{subeqnarray}}

\def\a{\alpha}
\def\b{\beta}
\def\g{\gamma}
\def\c{\chi}
\def\d{\delta}
\def\e{\epsilon}

\def\f{\phi}

\def\k{\kappa}
\def\l{\lambda}
\def\m{\mu}
\def\n{\nu}
\def\o{\omega}

\def\th{\theta}

\def\r{\rho}

\def\t{\tau}

\def\x{\xi}

\def\cA{{\cal A}}
\def\cF{{\cal F}}

\def\cN{{\cal N}}

\begin{document}


\begin{titlepage}

\title{
   \hfill{\ns Imperial/TP/02--03/1\\}
   \hfill{\ns hep-th/0210228\\}
   \hfill{\ns \\[2cm]}
   {\LARGE $D=5$ M-theory radion supermultiplet dynamics}}

\author{
   Jean-Luc Lehners\footnote{Email: jean-luc.lehners@ic.ac.uk} and K.S. Stelle\footnote{
Email: k.stelle@ic.ac.uk}\\[0.5cm]
   {\ns The Blackett Laboratory, Imperial College, London SW7 2BW, UK}}

\date{}

\maketitle

\begin{abstract}\ns
We show how the bosonic sector of the radion supermultiplet plus $d=4$, $\cN=1$
supergravity emerge from a consistent braneworld Kaluza-Klein reduction of $D=5$
M--theory. The radion and its associated pseudoscalar form an
$SL(2,\mathbb{R})/U(1)$ nonlinear sigma model. This braneworld system admits its
own brane solution in the form of a 2-supercharge supersymmetric string. Requiring
this to be free of singularities leads to an $SL(2,\mathbb{Z})$ identification of
the sigma model target space. The resulting radion mode has a minimum length; we
suggest that this could be used to avoid the occurrence of singularities in
brane-brane collisions. We discuss possible supersymmetric potentials for the
radion supermultiplet and their relation to cosmological models such as the cyclic universe or hybrid
inflation.
\end{abstract}

\thispagestyle{empty}

\end{titlepage}


\section{Introduction}


Heterotic string theory is one of the most promising string
theories for phenomenological applications. In particular, models
resembling the Standard Model can be obtained as low energy
solutions of this theory. It is therefore not surprising that
cosmological models inspired by heterotic string theory have
recently been tried out. The proper setting for such theories lies
within heterotic M--theory: Ho\v{r}ava and Witten showed how
heterotic string theory arises from M--theory by compactifying
11--dimensional supergravity on the orbifold
$S_{1}/\mathbb{Z}_{2}$, and by including boundary theories with a
gauge group $E_{8}$ on each of the two orbifold fixed points in
order to cancel gravitational anomalies \cite{HW}. For fields
slowly varying across the orbifold this construction has as its
low energy limit heterotic string theory with gauge group
$E_{8}\times E_{8}$ in ten dimensions. One can then compactify
this on a Calabi--Yau manifold in order to retrieve 4--dimensional
physics. However it turns out that for phenomenological reasons
the size of the orbifold has to be about an order of magnitude
larger than the Calabi--Yau size \cite{BD}. Hence going up in
energy from an initial 4--dimensional point of view, the world
would first look 5--dimensional and then 11--dimensional. Moreover
the vacuum of the 5--dimensional theory is not flat space, but has
been shown to consist of two parallel and static 3--branes located
at the orbifold fixed points \cite{LOSW}. One may then try to
identify one of the 3--branes with our visible universe.
Cosmological scenarios, such as the cyclic universe of Steinhardt
and Turok \cite{ST}, for example, (or its predecessor, the
ekpyrotic universe) \cite{KOST}, or the models of Brax and Davis
\cite{BXD}, obtain an expanding universe on the brane worldvolume
as a consequence of relative motion of the branes. Relative motion
of the initially static branes can be achieved via the inclusion
of a conjectured non--perturbative potential for the radion field
determining the distance between the branes.

In this article, we will attempt to clarify certain mathematical issues concerning
this setup: first of all we will identify the pseudoscalar partner
of the radion in an $\cN=1$ chiral supermultiplet. We then
show that it is consistent to truncate the original 5--dimensional
theory down to the 4--dimensional worldvolume of the brane while
keeping $d=4$ gravity and the radion supermultiplet scalars, {\it i.e.}\ the
5--dimensional equations of motion reduce to 4--dimensional ones that are
independent of the orbifold direction. The radion supermultiplet scalars form an
$SL(2,\mathbb{R})/U(1)$ nonlinear sigma model. On the 4--dimensional worldvolume, we
will construct a solitonic string solution supported by this sigma model, in which
the pseudoscalar may be viewed as a 0--form gauge potential. This
string has finite energy only if we one makes identifications in the target space
under a discrete $SL(2,\mathbb{Z})$ subgroup of $SL(2,\mathbb{R})$, so that the
reduced target space becomes $SL(2,\mathbb{Z})\backslash SL(2,\mathbb{R})/U(1)$
\cite{GSVY,GGP,BC}. In fact, $SL(2,\mathbb{Z})$ has been conjectured to be preserved
as a local symmetry in this sense of the full quantum string theory \cite{SS}. This
has interesting consequences, especially if we look at the string solution from
a 5--dimensional point of view. Indeed, in 5 dimensions the
string arises as the intersection of a membrane with the boundary
3--branes, as we will show. And the $SL(2,\mathbb{Z})$ identification of the target
space implies that the distance between the boundary branes has a minimum value, a
result of potential significance for cosmological models relying on the collision
of boundary branes.

In order to clarify the r\^{o}le of the pseudoscalar further,
we will look at the conditions under which it itself can be
truncated out. In fact, the pseudoscalar may be ignored entirely in a
consistent way in the absence of a potential. However the question of
whether one can truncate it must be reviewed when one includes a potential, since
such a potential generically leads to interactions between the scalar and
pseudoscalar. After these considerations we will show that the potential
proposed for the cyclic universe cannot be embedded in heterotic M--theory,
even if one neglects the $SL(2,\mathbb{Z})$ symmetry. It can, however,
be approximated to some extend. We will give an example of such a supersymmetric
approximation and will give the completion of it to a
two--field potential after reinstating the pseudoscalar.

Finally we will turn our attention to the corrections that arise
from integrating out the massive modes that occur in
Calabi--Yau compactifications, which can couple to the massless
sector. These corrections, in the presence of a superpotential,
turn out to affect not only the kinetic terms of the radion and
pseudoscalar, but also the potential of the theory. The
importance of these corrections will be studied for known
non-perturbative and flat potentials as well as a supersymmetric approximation to
the cyclic universe potential.

\section{A Consistent Truncation to Gravity and Scalars}

The bosonic sector of the bulk action for 5--dimensional heterotic M--theory,
including gravity and the universal hypermultiplet is given by
\cite{LOSW}
\bea
S_5 &=& \frac{1}{2\k_5^2}\int_{M_5}\sqrt{-g}\left[
                  R-\frac{3}{2}\cF_{\a\b}\cF^{\a\b}-\frac{1}{\sqrt{2}}
                 \e^{\a\b\g\d\e}\cA_\a\cF_{\b\g}\cF_{\d\e}-\frac{1}{2}V^{-2}\partial_\a V\partial^\a V
                   -2V^{-1}\partial_\a\x\partial^\a\bar{\x} \right.\nn \\
                   && \left.-\frac{1}{24}V^2G_{\a\b\g\d}G^{\a\b\g\d}-\frac{\sqrt{2}}{24}\e^{\a\b\g\d\e}G_{\a\b\g\d}
                   (i(\x\partial_\e\bar{\x}-\bar{\x}\partial_\e\x
                   )-2\a\cA_\e)-\frac{1}{3}V^{-2}\a^2\right]\ . \label{D5act}
\eea
The static 3--brane solution for this system is given by \cite{LOSW}
\bea
 ds_5^2 &=& H(y)dx^\m dx^\n\eta_{\m\n}+H^4(y)dy^2  \\
 V &=& H^3(y)\ ,
\eea
with the codimension one harmonic function
\be
H(y) = \frac{\sqrt{2}}{3} \a |y| +c_{0}\ ,
\ee
and all other fields set to zero. In order to give a $D=5$ Ho\v{r}ava--Witten
geometry, the harmonic function is taken to have a second kink at $y = \pi \r.$ The
solution then represents a pair of parallel 3--branes supported by the scalar $V$ at
coordinate positions $y=0,\pi \r.$ Note that the distance between branes is
given by $d =\int_{0}^{\pi \r}H^2 dy$, and is thus static but
arbitrary since the coordinate position $\pi \r$ may be chosen at
will.

We wish to extend the above solution to one in which the branes
can move relative to each other, {\it i.e.}\  in which the size of
the orbifold can vary. Introducing the radion $b(x^{\m})$, a
function of the worldvolume coordinates only, we state the
following ansatz: 
\be 
ds_5^2 = e^{-\frac{b}{2}}H(y) g_{\m\n} dx^\m
dx^\n + e^{b}H^4(y)dy^2. \label{metans} 
\ee 
This metric is a solution of the theory if 
\be 
V = e^{\frac{b}{2}} H^3\ . 
\ee 
Now the interbrane distance 
\be 
d=\int_{0}^{\pi \r} e^{\frac{b}{2}}H^2 \label{branedist}
dy 
\ee 
is a function of the worldvolume coordinates and is
dynamical. Note that inclusion of the modulus $b$ makes the
parameter $\rho$ redundant, as we may eliminate it by a rescaling
of $y$. So we shall set $\rho=1$ hereafter. The branes will
collide as $b \to - \infty$, and in this limit $V\to 0$. Since
$V$, from the 11--dimensional point of view, represents the volume
of the compactified Calabi--Yau space, we see that at the moment
of collision a total of $1+6=7$ dimensions disappear. Cosmological
theories in which the branes collide, {\it e.g.}\ the ekpyrotic
universe, need to take into consideration that the Calabi--Yau
dimensions also disappear momentarily. The way in which string
theory handles this singularity is not clear at present, and some
recent studies seem to indicate that such singularities may
persist even in the full theory \cite{LMS}.

The domain wall solution preserves 4--dimensional $\cN = 1$
supersymmetry on the brane worldvolume. Scalars therefore belong
to chiral supermultiplets, and the radion must be paired with a
pseudoscalar to form a complex scalar. Let us denote the
pseudoscalar by $\c(x^{\m})$. It can also be obtained from a consistent truncation
of the $D=5$ theory, given by setting
\bea
\cF_{\m
y} &=& H(y)^2 \pt_{\m}{\c} \label{Fans}\\
G_{y \m \n \r} &=& e^{-b} \e_{\m \n
\r}^{\l} \pt_{\l}{\c}\ ,\label{Gans}
\eea
where $\e_{\m \n \r \l}$ is the
4-dimensional Levi-Civita tensor, and all other unrelated index
structures of $\cF$ and $G$ as well as the complex scalar $\x$ are set to zero.
Taken together with $d=4$ gravity and the radion mode $b$, the 5--dimensional
equations of motion reduce to a 4--dimensional system that can be
summarized by the following effective action:
\be
S_{4} = \frac {1}{2 \k^2}
\int_{M_{4}} \sqrt{-g}(R^{(4)}-\frac{1}{2}\pt_{\r}{b}
\pt^{\r}{b}-4 e^{-b} \pt_{\r}{\c} \pt^{\r}{\c})\  . \label{d4act}
\ee
Note that, as a consequence of the consistency of this reduction, (\ref{d4act}) may
also be obtained directly by substituting the reduction ansatz
(\ref{metans},\ref{Fans},\ref{Gans}) into (\ref{D5act}), but the proper
verification of this consistency is made using the $D=5$ equations of
motion.\footnote{The possibility of consistently reducing from a surrounding bulk
supergravity theory to a braneworld supergravity multiplet was found in Refs
\cite{bonb}, but such consistent reductions do not in general allow for
the retention of matter multiplets. Hence the consistency of the $d=4$
supergravity/radion supermultiplet system was not {\it a priori} expected.}

Since the theory is supersymmetric it should be possible to cast
the kinetic terms in K\"{a}hler form. For this purpose let us
define a complex scalar
\be
\f \equiv e^{\frac{b}{2}} + i \sqrt{2}
\c\ .
\ee
Then the K\"{a}hler potential can be identified to be
\be
K = -4 \ln{(\f + \bar{\f})}\ .\label{kahlerpot}
\ee
From this, or more directly from
the 4--dimensional effective action, it can be seen by inspection
that the sigma model formed by the two scalars $b$ and $\c$ has a
$SL(2,\mathbb{R})$ symmetry. Also, in the notation of Ref.\ 
\cite{LOSW}, our consistent truncation turns out to be the
combination $\f = S = T.$

\section{The Solitonic String solution}

Our 4--dimensional bosonic theory consists of gravity and two scalars. The
exponential coupling of the second scalar to the first indicates
that this theory should have a brane solution. The scalar $\c$
acts as a 0--form gauge field, giving rise to a one--form field strength.
In 4 dimensions, this is dual to a three--form field strength,
which couples to a string solution. Explicitly, we have
\bea
ds_4^2 &=& \eta_{\m\n}dx^\m dx^\n+h^4(r)dx^m dx^m \    \ \m=0,1 \
\ m=2,3 \\
e^{b} &=& h^2(r)  \\
\sqrt{2}\c_{,m} &=&
\e_{mn}h(r)_{,n}
\eea
with the harmonic function $h(r)=1+ln(r)$
depending on the radial transverse distance $r=\sqrt{x^m x^m}$.
The solitonic string can be compared to a cosmic string: the
spacetime is conical, with a singularity at the location of the
string ($r=0$). However this solitonic string possesses
non--vanishing energy--momentum throughout space, unlike the
cosmic string. Note that in this solution the metric is also
degenerate at $r=e^{-1}$, where there is a curvature singularity and infinite
energy density. We shall shortly see how this is to be avoided; the above behavior
will however remain valid at large $r$.

In order to find a better behaved solution, let us introduce the
complex field 
\be 
\t \equiv \t_{1}+i \t_{2} \equiv \sqrt{2}\c+i
e^{\frac{b}{2}}=i \bar{\f}\ . 
\ee 
The 4--dimensional action can then
be rewritten as 
\be 
S_{4} = \frac {1}{2 \k^2} \int_{M_{4}}
\sqrt{-g}(R^{(4)} - 2 \frac{(\nabla \t_1)^2 +(\nabla
\t_2)^2}{\t_{2}^2})\ . \label{tauact} 
\ee 
Now assume that $\t$ here depends only on $x^2$ and $x^3$, since we are looking for a
string solution. Writing $\pt=\frac{\pt}{\pt z}$ and $z=x^2 +i x^3$, the equation of
motion for $\t$ is given by 
\be 
\pt\overline{\pt} \t + 2 \frac{\pt \t \overline{\pt}
\t}{\overline{\t} - \t}=0\ . 
\ee 
It is solved by any holomorphic
$\t = \t(z).$ Our metric ansatz reads 
\be 
ds_4^2 = -dt^2 + dx^2
+E^4(z,\bar{z})dz d\bar{z}\ , 
\ee 
with the harmonic function given by 
\be 
E(z,\bar{z}) = \frac{1}{2} [h(z) + \bar{h}(\bar{z})] 
\ee
where $h(z)$ is a holomorphic function. The Einstein equations are
solved for 
\be 
\t(z) = i h(z)\ . 
\ee 
It turns out \cite{GSVY,GGP,BC} that in order to have finite energy solutions
one must make sigma model target-space identifications under an
$SL(2,\mathbb{Z})$ subgroup of the $SL(2,\mathbb{R})$ symmetry of
(\ref{tauact}). $\t$ then takes its values in the upper half
complex plane modulo $SL(2,\mathbb{Z})$ transformations, {\it
i.e.}\ the moduli space becomes $SL(2,\mathbb{Z})\backslash
SL(2,\mathbb{R})/U(1)$ (note that this makes $\c$ periodic).
Physically inequivalent values of $\t$ lie in the fundamental
domain $F$ of the modular group, defined by $-\frac{1}{2} \le
\t_{1} < \frac{1}{2}$ and $|{\t}| \ge 1$. There exists a map,
called the $j$--function, which maps $F$ in a one--to--one and
holomorphic fashion onto the complex plane \cite{GSVY}. Hence we
can specify a holomorphic function and pull it back to $F$ using
$j$, thus obtaining a solution for $\t$ which respects the
$SL(2,\mathbb{Z})$ symmetry. A single string solution is given by
\be 
j(\t(z)) = z\ . 
\ee 
Multiple string solutions are obtained by
taking $j(\t)$ to be a polynomial in $z$. Asymptotically we have
$\t \sim \frac{i}{2 \pi} \ln(z),$ so that $h(z) \sim ln(z)$ and
$E(z,\bar{z}) \sim \ln(|z|)$ for large $|z|.$ This means that we
recover a cylindrical symmetry, and actually have a conical
spacetime at locations far away from the string.

Let us also check the supersymmetry conditions that this solution satisfies. These
are derived from the requirement that one be able to set the fermionic fields to
zero consistently with the surviving supersymmetry. The transformations of the
spinor partner of
$\t$ and of the gravitino lead to the following requirements:
\bea
\delta
\lambda &=& 0 \  \ \Rightarrow \   \ \g^{a} \pt_{a} \t \bar{\e} = 0 \label{deltalambda}\\
\delta \psi_{\m} &=& 0 \  \ \Rightarrow \   \ \pt_{\m} \e +
\frac{1}{4} \o_{\m}^{a b} \g_{a} \g_{b} \e - \frac{\pt_{\m}(\t +
\bar{\t})}{\t - \bar{\t}} \e = 0\ ,\label{deltapsi}
\eea
where $\o_{\m}^{a b}$ is
the spin connection and $\e$ is a $d=4$ Majorana spinor parameter. Now using $\g_z =
\frac{1}{2} (\g_2 - i \g_3)$ and $\g_{\bar{z}} = \frac{1}{2} (\g_2
+ i \g_3)$ we obtain
\be
\frac{1}{4} \o_{\m}^{a b} \g_{a} \g_{b} =
\frac{\pt_{\m}(\t + \bar{\t})}{\t - \bar{\t}} i \g_{23}\ , 
\ee
with $\g_{23} = \frac{1}{2} (\g_2 \g_3 - \g_3 \g_2)$. Imposing
the condition 
\be 
\g_{23} \e = - i \e
\ee
then solves equation
(\ref{deltapsi}). This condition is equivalent to demanding that
\be
P \e
\equiv \frac{1}{2} (1 - i \g_{23}) \e = 0\ ,
\ee
where $P$ is a
projection operator satisfying $P^2 = P.$ We have thereby imposed a $d=2$ worldsheet
chirality constraint on the spinor parameter $\e$ which projects
out half of its components. We are thus left with just 2 supercharges
as expected, thus preserving half of the 4--dimensional $\cN = 1$
supersymmetry. Equation (\ref{deltalambda}) is solved by virtue of the projection
condition and holomorphicity of the field $\t.$

\begin{figure}[ht]
\begin{center}
\includegraphics{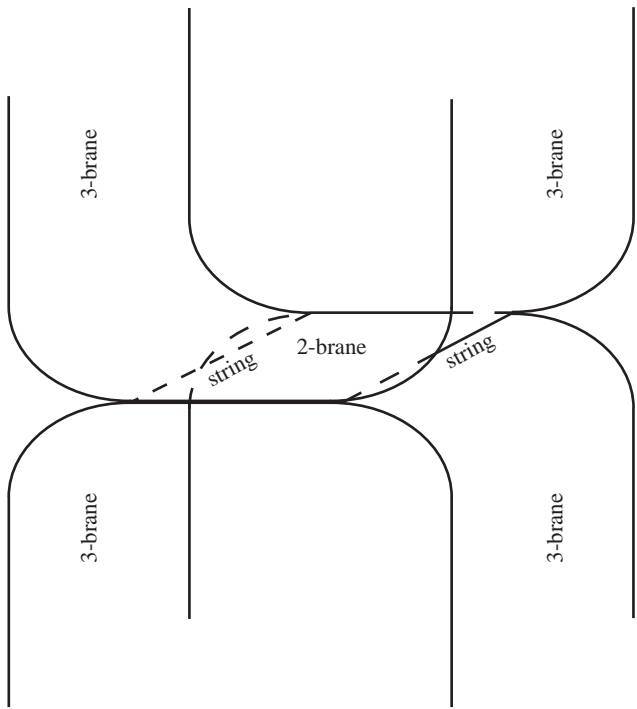} \caption{The
five-dimensional interpretation of the solitonic string as the
intersection of a membrane with the two boundary 3-branes, with the intersecting
string delocalized along the membrane.}\label{fig1}
\end{center}
\end{figure}

Since we have a consistent truncation from $D=5$ to $d=4$, the
solitonic string solution can be oxidized straightforwardly back up to 5
dimensions, giving the resulting metric
\be
ds_5^2 =
\frac{H(y)}{E(z,\bar{z})}[-dt^2+dx^2+E^4(z,\bar{z})dz
d\bar{z}]+E^2(z,\bar{z}) H^4(y)dy^2\ .
\ee
In order to interpret
this solution \cite{G}, we should rewrite it as
\be
ds_5^2 =
\frac{H}{E_1}(-dt^2+dx^2)+H E_1^2 E_2 dz d\bar{z}+E_1^2 H^4 dy^2\ ,
\ee
with $E_1=E_2=E(z,\bar{z}).$ Now taking limits in which we
set the harmonic functions $E_1, E_2$ and $H$ pairwise equal to
$1,$ we recover the metrics of the 3--brane ($E_1=E_2=1$), a
membrane ($E_1=H=1$) and a string ($E_2=H=1$) in 5 dimensions,
with the string being delocalized in the y--direction.\footnote{Such semi-localized
intersecting brane solutions have been discussed, {\it e.g.}\ in Ref.\ \cite{peet}.}
Comparing with the full 5--dimensional action (\ref{D5act}) we can verify that the
membrane is supported electrically by $G_{\a \b
\g
\d}$ and the string is supported magnetically by $F_{\a \b}.$ Thus our
5--dimensional interpretation of the solitonic string solution is as an
intersection of one 3--brane and a 2--brane, which then stretches between
it and the other orbifold 3--brane, and with the intersection string delocalized
over the worldvolume of the 2--brane. This setup is sketched in Figure \ref{fig1}.
Note that we have
\be
\t_2 = e^{\frac{b}{2}} = \frac{1}{2} [
h(z)+\bar{h}(\bar{z})] = E(z, \bar{z})\ .
\ee

We can draw two
observations from this equation, remembering that
$e^{\frac{b}{2}}$ governs the distance between the orbifold branes
({\it cf.}\  Equation (\ref{branedist})): {\it i)} the distance grows as
$\ln{|z|}$ for large $|z|$ and {\it ii)} since $\t$ takes its
values only in the fundamental domain F, there is a non--zero
minimal distance between the branes at which
\be
e^{\frac{b}{2}} =
\frac{\sqrt{3}}{2}\ .
\ee
Thus the reduction of the target space from
$SL(2,\mathbb{R})/U(1)$ to $SL(2,\mathbb{Z})\backslash SL(2,\mathbb{R})/U(1)$
prevents the collision of the two boundary branes. This is of significance since
the full quantum string theory is expected to exhibit such a local
$SL(2,\mathbb{Z})$ symmetry \cite{SS}. In this light, cosmological scenarios
relying on a collision between the boundary branes would need to be revised.
Moreover, one might speculate that this reduction of the target space could provide
a mechanism for the two branes to bounce after reaching their
minimum separation in the presence of an attractive potential.\footnote{The scale
of the minimum separation would be set by the tension of the 3-brane worldvolume
string, similarly to the way in which $\alpha'=1/(2\pi T)$ sets the minimum length
$\sqrt{2\pi\alpha'}$ as a consequence of T--duality, under which $r\rightarrow
2\pi\alpha'/r$.}

However, for the rest of this paper, we will proceed by considering the case of an
unreduced target space $SL(2,\mathbb{R})/U(1)$.

\section{Truncating the Pseudoscalar}

The pseudoscalar $\c$ can be set to zero consistently if the
equations of motion for the metric and $b$ can still be satisfied
at all times when $\c = 0$. In the absence of a potential, this is
possible since $\c$ does not have any $\c$--independent source
terms. Then the radion $b$ becomes a free massless scalar
satisfying $\Box^{(4)}{b} = 0$. But a potential $V(\f,\bar{\f})$
is, in general, a function of $b$ and $\c$. And if it contains
terms of the form $\c f(b)$ ,where $f(b)$ is an arbitrary function
of $b$, then the resulting equation of motion for $\c$ will be 
\be
D^{(2)} \c = f(b)\ , 
\ee 
where $D^{(2)}$ is a second order
differential operator. Clearly in this case it is inconsistent to
neglect $\c$. Thus we can conclude that the condition for the
truncatability of $\c$ reads 
\be 
\frac{\pt V}{\pt
\c}\arrowvert_{\c = 0} = 0\ . 
\ee

Any potential $V(\f,\bar{\f})$ in supergravity arises from a
superpotential $W(\f)$ (which is a holomorphic function of $\f$)
by the formula 
\be 
V = e^{K} [K^{\f \bar{\f}}D_{\f}W
\overline{D_{\f}W} - 3 W \overline{W}] \label{VWrel}
\ee 
with $D_{\f}=\pt_{\f}+\frac{\pt K}{\pt \f}$, where $K$ is the
K\"{a}hler potential and $K^{\f \bar{\f}}=(\frac{\pt^2{
K}}{\pt{\f} \pt{\bar{\f}}})^{-1}$. $K$ was given above in equation
(\ref{kahlerpot}) and so we obtain in the present theory 
\be
V(\f,\bar{\f})=\frac{1}{4(\f+\bar{\f})^{2}} \frac{\pt W}{\pt
\f}\frac{\pt \overline{W}}{\pt \bar{\f}} -
\frac{1}{(\f+\bar{\f})^{3}} (\frac{\pt W}{\pt \f} \overline{W} + W
\frac{\pt \overline{W}}{\pt \bar{\f}}) +
\frac{1}{(\f+\bar{\f})^{4}} W \overline{W}\ . 
\ee 
Thus we have
\bea 
\frac{\pt V}{\pt \c} &=&
\frac{i\sqrt{2}}{4(\f+\bar{\f})^{2}}(W,_{\f\f}\overline{W},_{\bar{\f}}
- W,_{\f} \overline{W},_{\bar{\f}\bar{\f}})\nn \\
&&- \frac{i\sqrt{2}}{(\f+\bar{\f})^{3}}(W,_{\f\f}\overline{W} - W
\overline{W},_{\bar{\f}\bar{\f}})\nn \\
&&+ \frac{i\sqrt{2}}{(\f+\bar{\f})^{4}}(W,_{\f}\overline{W} - W
\overline{W},_{\bar{\f}})\ .
\eea
Setting $\c=0$, we have
$\f=\bar{\f}$. Hence we see that the condition for the
truncatability of $\c$ becomes (as can be verified using a series
expansion of $W$)
\be
W = \overline{W}\quad {\rm for}\quad \phi=\bar\phi\ ,
\ee
up to an irrelevant
overall phase. Thus, if the superpotential is real when $\c$ is set to
zero, then $\c$ can be truncated and the formula for the potential
becomes
\be
V = \frac{1}{16 \f^{2}} (W,{\f}-\frac{2W}{\f})^{2} -
\frac{3}{16} \frac{W^{2}}{\f^{4}}\ .\label{uncorrpot}
\ee
But for a general complex superpotential we are not allowed to disregard $\c$ and it
may not be dynamically sensible to do so even if it is mathematically consistent.
This could be important for cosmology. We in general have two interacting
scalars and both scalars need to be taken into account when studying the
consequences of inflation, density perturbations or quintessence.

For future reference, let us pause here to calculate the solutions
to (\ref{uncorrpot}) leading to a flat potential $V=V_0$ representing a
cosmological constant. Setting $\tilde{W} = \f^{-2} W$ simplifies
the equation, since we can rewrite it as
\be
\f^2 \tilde{W}_{,
\f}^2 = 3 \tilde{W}^2 + 16 V_{0}\ .
\ee
For $V_{0} < 0$, the
solutions are
\bea
W &=& 4 \sqrt{- \frac{V_{0}}{3}} \f^2 \label{1stsuppot}\\
W &=& 2 \sqrt{\frac{-V_{0}}{3}}(\f^{2 + \sqrt{3}} + \f^{2 -
\sqrt{3}})\ , \label{2ndssuppot}
\eea
whereas for $V_{0} > 0$ the solution is
\be
W = 2
\sqrt{\frac{V_{0}}{3}}(\f^{2 + \sqrt{3}} - \f^{2 - \sqrt{3}})\ . \label{posVsuppot}
\ee

\section{A Note on the Cyclic Universe Potential}

The potential proposed in the cyclic universe scenario \cite{ST}
is of the form
\be
V(\f)
 = V_{0} e^{-\frac{1}{\f}}(1 - \f^{-10})\ . \label{cyclicpot}
\ee
The radion is the
only scalar being considered: $\c$ has been set to zero and accordingly
$\f = e^{\frac{b}{2}}$. A serious criticism of this conjectured potential is that
it cannot in fact be derived from a superpotential,
and therefore cannot exist in heterotic M--theory. For its
existence one would need to solve
\be
16 V_{0}
 e^{-\frac{1}{\f}}(1 - \f^{-10}) = (\frac{W,{\f}}{\f}-\frac{2W}{\f^{2}})^{2} -
\frac{3W^{2}}{\f^{4}}\ .
\ee
To see why this cannot be solved for $W$, given $V$ in (\ref{cyclicpot}), let us
try to solve this equation as
$\f \to 0$. Then the dominant term on the left is $ - 16 V_{0}
e^{-\frac{1}{\f}} \f^{-10}$ and because it is negative we
require on the right that the negative second term be dominant in this
limit. This gives $W \to 4 \sqrt{\frac{V_{0}}{3}} e^{-\frac{1}{2
\f}} \f^{-3}$ as $\f \to 0.$ But for this $W$ it is in fact the first
(positive) term on the right that dominates, in
contradiction with our assumptions. A last possibility would be
that both terms on the right are equally dominant. This would be
the case if $W$ were a power of $\f$. But a polynomial $W$ cannot
reproduce the non-perturbative $e^{-\frac{1}{\f}}$ factor in
(\ref{cyclicpot}). Note that this also applies for any negative power of $\f$
in (\ref{cyclicpot}); the power $-10$ is of no special importance for this argument. We
conclude that the cyclic universe potential cannot exist in heterotic M--theory.

\begin{figure}[ht]
\begin{center}
\includegraphics{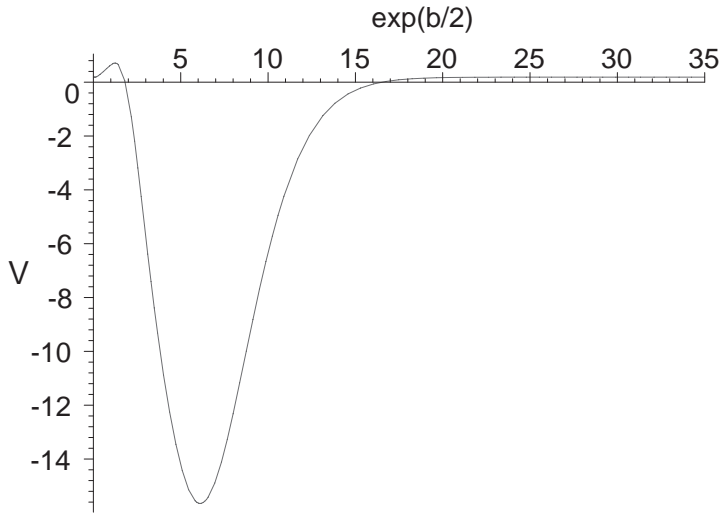}
\caption{The approximation to the cyclic universe potential
necessarily exhibits a positive bump near the origin.}\label{fig2}
\end{center}
\end{figure}

However, to some extent the potential (\ref{cyclicpot}) can be
supersymmetrically approximated, and in particular for the region
in $e^{\frac{b}{2}}$ that is relevant for calculating the spectrum of density
fluctuations. (We note also that in this region the cyclic
potential coincides closely with the ekpyrotic one.) We take the
approximate superpotential to be 
\be 
W(\f) = e^{-\frac{1}{2 \f}}c
\f^5 e^{-\f} + W_{flat}\ , 
\ee 
where $W_{flat}$ denotes the flat
solution found above in (\ref{posVsuppot}). This gives a potential
of the shape plotted in Figure \ref{fig2}; $c$ determines the
depth of the dip. The shape is similar to that of the cyclic
potential (\ref{cyclicpot}) except for the (inevitable) little
positive bump near the origin. Now the branes can only collide if
the scalar field picks up enough kinetic energy to overcome this
bump.

\begin{figure}[ht]
\begin{center}
\includegraphics{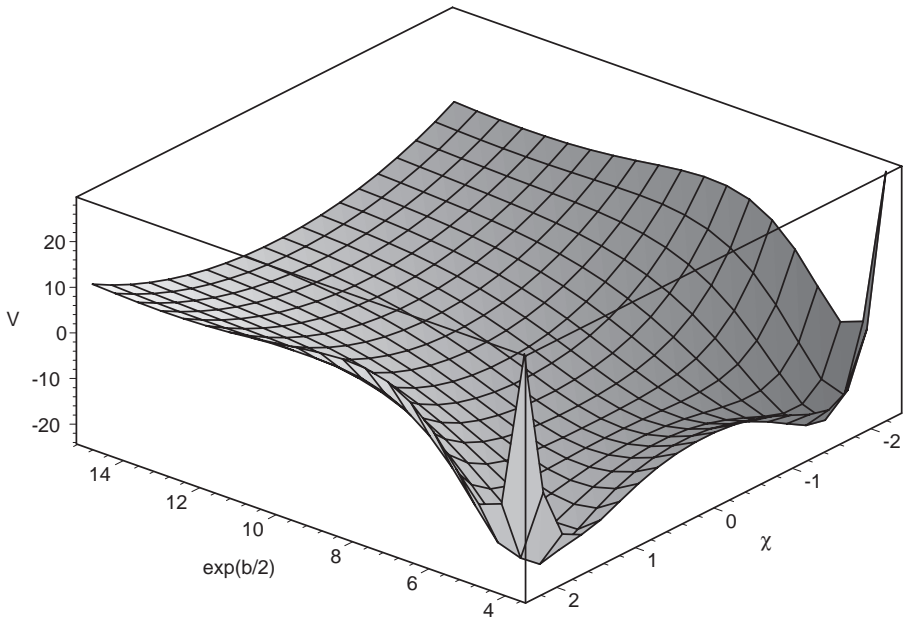}
\caption{The part of the two-field potential that is relevant for a calculation
of the spectrum of density perturbations. The $\c=0$ section
represents the original one-field potential. Note the instability
in the form of a saddle point at the minimum of the $\c=0$
section.}\label{fig3}
\end{center}
\end{figure}

We can restore the second scalar and plot the resulting two--field potential in
the steep region where the density fluctuations are evaluated (see Figure \ref{fig3}). Note
that, as one comes in from large $b$ values, the potential falls
in the $e^{\frac{b}{2}}$ direction, but the transverse curvature
becomes larger and larger, thus tending to confine the scalar
field to the $\c=0$ plane. However, there is an instability at the
minimum of the original one-field potential: at the saddle point,
the field will generically roll off to one of the true minima of the
potential where $\c$ is equal to plus or minus a non-zero constant. It
might be interesting to see what implications this has for
cosmology, bearing in mind that in this model the adiabatic modes
of density fluctuations will be complemented by isocurvature
modes.

{\it En passant,} we note that the shape of the potential shown in Figure \ref{fig3} is reminiscent of the
potentials considered in hybrid inflation scenarios \cite{hybrid}, so the cosmological applications of the
radion dynamics considered in this paper could perhaps also be extended to inflationary models.

\section{Kaluza--Klein Corrections to Potentials}

In the Kaluza--Klein reduction from eleven to five
dimensions on a Calabi--Yau manifold, there can be couplings of
massive graviton supermultiplets (the most significant of which will have a mass
roughly equal to the inverse radius of the Calabi--Yau manifold) to the
massless scalar $\f$, couplings which can be linear in the
massive supermultiplet fields. Not enough is known about Calabi--Yau manifolds to
enable a detailed calculation of these couplings, so this should be considered as an
indicative study of eventual couplings of this type. Given a coupling linear in the
massive graviton supermultiplet fields, the massive modes cannot be ignored and must
be properly integrated out, as discussed in Ref.\ \cite{DFPS}, the relevant results
of which we now review.

The analysis is done in superspace. The radion and its pseudoscalar partner $\c$ plus
their fermionic superpartner $\lambda$ together with the associated auxiliary fields are
component fields of a chiral superfield that we shall also denote by $\phi$. When we come
to compare the forms of corrected and uncorrected potentials, we can set $\c=\lambda=0$
and eliminate the auxiliary fields, taking once again $\f = e^{\frac{b}{2}}$; but for now
we shall remain in superspace. Since we consider here the coupling problem to lowest order
in a massive graviton supermultiplet, it is sufficient to restrict attention to the
linearized level. Linearized supergravity is represented by a vector superfield $V_{\a
\dot{\a}}$, and in order to be able to write an action invariant under both linearized
super--Poincar\'{e} and super--Weyl transformations, one must also include a chiral
conformal compensating multiplet, represented here by a prepotential superfield $X$. The
mass term for this supermultiplet is then given by
\be
\frac{m^{2}}{2} \int{d^{4}x d^{4}\th V_{\a \dot{\a}} V^{\a
\dot{\a}}} - \frac{m^{2}}{2} \int{d^{4}x d^{4}\th X \overline{X}}\ ;
\label{gravmass}
\ee
this combination is needed to reproduce the Pauli-Fierz mass term for a
ghost-free massive spin two field. Kaluza-Klein couplings of the massive graviton
supermultiplet to massless fields must be gauge--invariant in form (since gauge
invariance is broken only in the mass term), and therefore the massive graviton
supermultiplet must couple to a supercurrent
$J_{\a
\dot{\a}}$. The available supercurrent must be constructed out of the massless
scalar superfield $\f$ and gauge invariance implies that it must obey the
conservation law
\be
D^{\a} J_{\a \dot{\a}} = \overline{D_{\dot{\a}}} \overline{S}\ ,
\ee
where
\be
\overline{S} = - \frac{1}{2} D^{2} K + 3
\overline{W}\ .
\ee
Here $K$ is the K\"{a}hler potential as given before
and $W$ is the posited superpotential. The kinetic action for the massless
scalars is given by the superspace integral of the K\"{a}hler
potential, {\it i.e.}\
\be
\int{d^4 x d^4 \th K}\ .
\ee
The non-truncatable couplings to the massive supermultiplet can be written as
\be
\int{d^{4}x d^{4} \th V_{\a \dot{\a}} J^{\a \dot{\a}}} -
\int{d^{4}x d^{2} \th (X S + \overline{X} \overline{S})}\ . \label{cplng}
\ee
Upon integrating out $V_{\a \dot{\a}}$ and $X$ while keeping only lowest
order correction terms, we may drop the kinetic action terms for
the massive multiplet (which we have not written out here) and
retain only the mass term (\ref{gravmass}) and the coupling term (\ref{cplng}).
All the correction terms so obtained are of higher derivative structure except for
\be
\frac{18}{m^2} \int{d^4 x d^4 \th W \overline{W}}\ ,
\ee
which
modifies the K\"{a}hler potential by
\be
K \to K_{corr} = K +
\frac{18}{m^2} W \overline{W}\ .
\ee
Thus, after higher Kaluza-Klein corrections are taken into account, the presence of
a superpotential modifies the kinetic structure of the massless
scalars in the theory. The significance of this correction depends
very much on the scale of compactification and on the shape of the
superpotential considered.

Recall from equation (\ref{VWrel}) that the potential $V$ depends on both
the superpotential $W$ and on the K\"{a}hler potential $K$. Now,
since $K$ has been modified, the potential $V$ acquires
corrections as well and we then have 
\be 
V_{corr} = e^{K +
\frac{18}{m^2} W \overline{W}} [(K + \frac{18}{m^2} W
\overline{W})^{\f \bar{\f}}D_{\f}W \overline{D_{\bar{\f}}W} - 3 W
\overline{W}]\ , 
\ee 
with the covariant derivative
$D_{\f}=\pt_{\f}+\frac{\pt K}{\pt \f} + \frac{18}{m^2} \frac{\pt
W}{\pt \f} \overline{W}$. However, since the correction to $K$ is
of the form $W \overline{W}$ this in fact does not change the
condition on truncatability of the pseudoscalar $\c$ ({\it i.e.}\
that $W$ should be real when $\phi=\bar\phi$), so we can still set
$\c$ to zero consistently if we assume that $W$ has real expansion
coefficients. This gives us the following expression for the
potential: 
\be 
V_{corr} = (2 \f)^{-4} e^{\frac{18}{m^2}W^2} \{
(\frac{1}{\f^2} + \frac{18}{m^2} W_{, \f}^2)^{-1}[W_{,
\f}(1+\frac{18}{m^2}W^2) - 2 \frac{W}{\f}]^2 - 3 W^2 \}\ . \label{Vcorr}
\ee

Superpotentials of non--perturbative origin have been investigated
in this context by Lima {\it et al.} \cite{LOPR} and Moore {\it
et al.} \cite{MPS}. They arise from membrane instantons and are
generically of the form $W = e^{- \f}$, when no additional matter
is considered on the brane worldvolume. They lead to an
exponentially falling potential $V$ which therefore acts as a
repulsive force between the branes. Moreover this potential gets
corrected very little by the massive Kaluza--Klein modes.

\begin{figure}[ht]
\begin{center}
\includegraphics{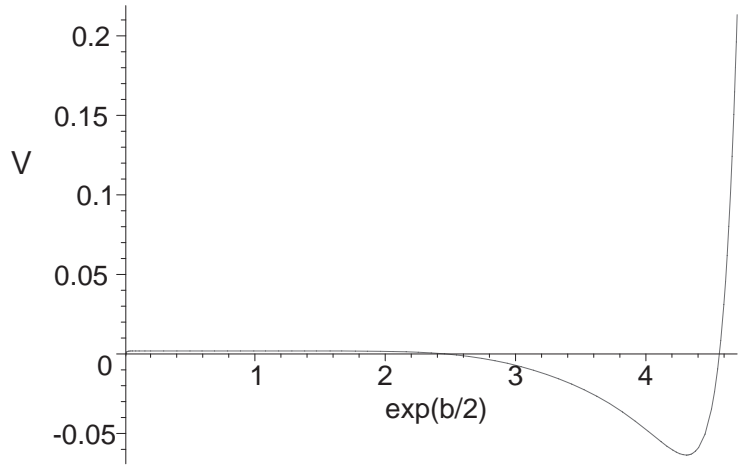}
\caption{Graph of an originally constant cosmological
potential after inclusion of massive Kaluza-Klein supermultiplet corrections. The
height $V_0$ of the original uncorrected potential determines the range of
$\f=e^{\frac{b}{2}}$ values where the corrections eventually become
important.}\label{fig4}
\end{center}
\end{figure}

More dramatic effects can be observed for potentials whose uncorrected forms do not
go to zero as $\f\rightarrow\infty$. Indeed, let us see
what the fate of a constant potential is when the corrections are
taken into account. We should note that it is not clear how such
potentials could be generated in the underlying microscopic theory, but many cosmological
models can be approximated in this way, at least asymptotically,
and so they are of interest to consider. A flat
potential is obtained by considering one of the superpotentials
(\ref{1stsuppot}--\ref{posVsuppot}). Now when we look at what the Kaluza--Klein
corrections do to this initially flat potential, the effect is
eventually drastic. As we go to large values of $\f$, we notice an
initial dip and then an exponential rise (see Figure \ref{fig4}). This is
due to the $e^{\frac{18}{m^2} W^2}$ prefactor in formula (\ref{Vcorr}).
This behavior is generic for potentials that do not tend to zero at
large values of $\f$. However, carrying out the same analysis for,
{\it e.g.,} the cyclic universe potential, which is
asymptotically flat, one notes that when plugging in
phenomenologically reasonable values of the parameters involved,
the Kaluza--Klein corrections only become important at values of
$\f$ so large that the corrections can be ignored in the region of cosmological
interest.

\section{Conclusions}

We have investigated the r\^{o}le of the chiral radion supermultiplet
in heterotic M--theory. The existence of a consistent truncation
from 5 to 4 dimensions is of central importance in our work. This
opens up the possibility of constructing cosmological scenarios
with dynamical boundary branes.

A prominent r\^ole in this paper is played by the pseudoscalar $\c.$
We showed that it cannot be ignored unless the superpotential can
be written as an expansion with only real coefficients, and even then ignoring it
may not be dynamically justified. In cases where it can be truncated we
estimated the effect of Kaluza--Klein corrections due to heavy modes on the
potential of the theory. They can be safely ignored for all potentials that
approach zero fast enough asymptotically for large radion values; otherwise they
tend to make the potential rise exponentially for large interbrane
distances, thus effectively putting an upper bound on the distance
between the boundary branes.

We constructed a solitonic string solution in $d=4$ dimensions
that is supported by the pseudoscalar $\c$. From the standpoint of
5 dimensions, the string appears as the intersection of a membrane
with a boundary brane. An important ingredient in consideration of
the solitonic string is the assumption that the full quantum
theory should exhibit an unbroken $SL(2,\mathbb{Z})$ symmetry that
is interpreted locally. As a consequence, the distance between the
boundary branes cannot reach zero, and thus the singularity at
which the orbifold and Calabi--Yau volumes vanish is avoided.
Virtual exchanges of membranes between the boundary branes may
lead to forces between the branes. It would be interesting to
calculate a potential for these forces from a purely
5--dimensional perspective. This potential should of course
respect the $SL(2,\mathbb{Z})$ invariance \cite{FKLZ}. But $V$ is
given by \be 
V(\t, \bar{\t}) = e^{K} [K^{\t \bar{\t}}D_{\t}W
\overline{D_{\t}W} - 3 W \overline{W}] \nn 
\ee 
and we have $K(\t,\bar{\t})=-4 ln(\t - \bar{\t})$ if we write $K$ in terms of $\t.$
Let $M$ denote an $SL(2,\mathbb{Z})$ transformation: 
\be 
M \t\equiv \frac{a \t + b}{c \t + d}\ , \     \ ad - bc = 1\ . \nn 
\ee 
For invariance of the potential, we want $V(M \t, M \bar{\t})=V(\t,
\bar{\t}).$ Now 
\be 
K(M \t, M \bar{\t})=K(\t, \bar{\t}) + 4 ln(c
\t + d) + 4 ln(c \bar{\t} + d)\ . \nn 
\ee 
This puts the following condition on the superpotential $W(\t)$: 
\be 
W(M \t)=(c \t +
d)^{-4} W(\t)\ . \nn 
\ee 
Noting that Dedekind's $\eta$--function
satisfies 
\be 
\eta(M \t)^{24} = (c \t + d)^{12} \eta(\t)^{24}\ ,
\nn 
\ee 
we can see that, for example, taking $W(\t) \propto
\eta(\t)^{-8}$ (multiplied by an arbitrary modular invariant
function) will give us an $SL(2,\mathbb{Z})$ invariant potential.
One would like to see which such non--perturbative superpotentials
can arise in the theory, and what their implications would be on
the dynamics of the radion mode.

\section*{Acknowledgments}

We would like to thank Andrei Linde, Andr\'e Lukas, Dieter L\"ust and Dan Waldram for discussions.

\end{document}